\documentclass[aps,prd,10pt,twocolumn,superscriptaddress,nofootinbib,showkeys,showpacs,altaffilletter]{revtex4-1}

\usepackage{graphicx}
\usepackage{dcolumn}
\usepackage{amssymb}
\usepackage{amsmath}
\usepackage{amsfonts}
\usepackage{amsbsy}
\usepackage{color}
\usepackage{rotating}
\usepackage[english]{babel}

\newcommand{\be}{\begin{equation}}
\newcommand{\ee}{\end{equation}}
\newcommand{\bea}{\begin{eqnarray}}
\newcommand{\eea}{\end{eqnarray}}

\begin{document}

\title{{\bf Modelling spatial variations of the speed of light}}

\date{\today}

\author{Adam Balcerzak}
\affiliation{Institute of Physics, University of Szczecin, Wielkopolska 15, 70-451 Szczecin, Poland}
\affiliation{Copernicus Center for Interdisciplinary Studies, S{\l}awkowska 17, 31-016 Krak{\'o}w, Poland}
\author{Mariusz P. D\c{a}browski}
\affiliation{Institute of Physics, University of Szczecin, Wielkopolska 15, 70-451 Szczecin, Poland}
\affiliation{Copernicus Center for Interdisciplinary Studies, S{\l}awkowska 17, 31-016 Krak{\'o}w, Poland}
\affiliation{National Centre for Nuclear Research, Andrzeja So{\l}tana 7, 05-400 Otwock, Poland}
\author{Vincenzo Salzano}
\affiliation{Institute of Physics, University of Szczecin, Wielkopolska 15, 70-451 Szczecin, Poland}


\begin{abstract}
We extend a new method to measure possible variation of the speed of light by using Baryon Acoustic Oscillations and the Hubble function onto an inhomogeneous pressure model of the universe. The method relies on the fact that there is a simple relation between the angular diameter distance $(D_{A})$ maximum and the Hubble function $(H)$ evaluated at the same maximum-condition redshift, which includes the speed of light $c$. One limit of such a method was the assumption of the vanishing of spatial curvature (though, as it has been shown, a non-zero curvature has negligible effect). In this paper, apart from taking into account an inhomogeneity, we consider non-zero spatial curvature and calculate an exact relation between $D_{A}$ and $H$.  Our main result is the evaluation if current or future missions such as Square Kilometer Array (SKA) can be sensitive enough to detect any spatial variation of $c$ which can in principle be related to the recently observed spatial variation of the fine structure constant (an effect known as $\alpha$-dipole).
\end{abstract}

\keywords{Cosmology, Inhomogeneity, Baryon Acoustic Oscillations, Speed of light}

\pacs{$98.80-k,98.80.Es,98.80.Cq, 04.50.Kd$}


\maketitle

\section{Introduction}

Recent observations of the quasar spectral lines have shown that the fine structure constant $\alpha$ does not only allow time variations \cite{alpha}, but also spatial variations \cite{alphadipole}. These variations known as $\alpha-$dipole are reported to be of Right Ascension $R.A. = 17.4 \pm 0.9$ h and declination $\delta = -58^{\circ} \pm 9^{\circ}$ or $(l, b) = (320^{\circ}, -11^{\circ})$ in galactic coordinates. More observations \citep{agafonova2011,webb12,molaro2013} and new more accurate methods of analysis \citep{webb15,deMartino16} have not dissipated the puzzle. Because of the definition of $\alpha = e^2/\hbar c$, where $e$ is the electron charge, $\hbar$ the Planck constant, one may relate the changes in $\alpha$ with the changes in $c$ - the speed of light i.e. $\Delta \alpha / \alpha = - \Delta c / c$.

Apparently, $\alpha-$dipole is roughly aligned with other dipoles such as the dark flow dipole, detected by \cite{watkins09} at $(l, b) = (287^{\circ}, 8^{\circ})$ using peculiar velocity measurements from galaxies (and few groups of clusters), and by \citep{kashlinsky08A,kashlinsky08B,kashlinsky10} at $(l, b) = (296^{\circ}, 140^{\circ})$ using the imprinting of clusters of galaxies velocity in the cosmic microwave background (Sunyaev-Zeldovich signal); or the dark energy dipole found by \cite{mariano12} at $(l, b) = (309^{\circ}, -15^{\circ})$ using both quasars and type Ia supernovae. This may suggest some large-scale inhomogeneous (spherically symmetric) distribution of matter in the universe which could perhaps be explained by allowing an inhomogeneous (i.e. non-friedmannian) model of the universe. All these detections are, anyway, puzzling and debated: improvements in the distance estimators can mitigate the dark flow dipole from galaxy velocities, but still being significant at a $98\%$ level \citep{watkins14}; officially, the \textit{Planck} team does not find any statistical evidence for dark flow \citep{PlanckDarkFlow}, but \citep{atrio13,atrio15} still claim on it; while for what concerns the dark energy dipole the statistical usefulness of supernovae has been found out to be null by \citep{SalzanoBeltran15,Javanmardi15}.

In fact, there have been very many discussions of whether the universe is really homogeneous beginning from the old ``observational cosmology'' program of Ellis and collaborators \cite{EllisOld} up to some more recent suggestions about a void inhomogeneity \cite{Tolman} which could explain the phenomenon of dark energy without any appeal to an idea of vacuum energy. A strongly explored model allowing dipole distribution of matter density was Lema\^{i}tre-Tolman-Bondi model found already in the 1930s \cite{LTB}.

Very recently, a vivid discussion about the role of local nonlinearities of gravitational interaction as to give the contribution to a global (``averaged'') energy-momentum tensor have been initiated in seminal paper by Green and Wald \cite{GW} partially inspired by Ref. \cite{Zalaletdinov}. In this paper, it was suggested that the ``averaging'' procedure by Buchert \cite{BuchertAveraging} cannot give the contribution which is of dark energy (negative pressure) type on the global level. Instead, it was claimed that the average contribution from inhomogeneities could only be gravitational waves which act as radiation with positive pressure. This was strongly objected in Ref. \cite{Buchert11} which pointed out that the assumptions on which the theorems leading to such a claim were questionable or inappropriate. Some exact solutions of Einstein field equations have been discussed in this context \cite{GWexample, Szybka14,Visser16}, looking for examples or counterexamples of the theorem.

The above further motivates studies of inhomogeneous models of the universe. A complementary to LTB model which allows a dipole is the inhomogeneous pressure model of Stephani \cite{Stephani}. This general model was investigated towards exact solutions in Refs. \cite{JMP93,dabrowski95,Sussmann,chris00}. Some of these solutions were confronted with data \cite{dabrowski98,stelmach01,GSS,PRD13,off-center14} imposing relatively strict bounds on the inhomogeneity, though not eliminating it completely \cite{PRD15}. In other words, a small dipole of inhomogeneous pressure is still possible.

In this paper we explore such a possibility saying that inhomogeneity is perhaps the reason for $\alpha-$dipole, though through $c-$dipole, due to the definition of the fine structure constant. The varying speed of light (VSL) theories have been explored and are described in Refs. \cite{VSL_theory} (for a more general review see Refs. \cite{Magueijo2003,Uzan2011}). Here, we allow that they are also applicable to inhomogeneous models of the universe which is a novelty. In the discussion, we follow some results from our recent papers in which the new method of the measurement of the speed of light through baryon acoustic oscillations (BAO) and cosmic chronometers was proposed \cite{PRL15,PRD16}.

The paper is organised as follows. In section \ref{InhPres} we briefly describe the properties of inhomogeneous pressure models. In Section \ref{Angdiam} we derive the angular diameter test for these models, and in Section \ref{data} we test them with observational data (supenovae, BAO, CMB) in order to find possible variation of $c$ or inhomogeneity. In Section \ref{results} we give our results and conclusions.

\section{Inhomogenoeus pressure Stephani universe}
\label{InhPres}

The Stephani universe is an inhomogeneous perfect-fluid energy-momentum tensor conformally flat solution of the Einstein field equations with a general spherically symmetric metric given by \cite{Stephani,JMP93}
\be
\label{STMET}
ds^2= -c_0^2 \frac{a^2}{\dot{a}^2} \left[ \frac{ \left( \frac{V}{a} \right)^{\centerdot} }{\left(\frac{V}{a}\right)} \right]^2 dt^2~
+ \frac{a^2}{V^2} \left[dr^2+r^2 d\Omega^2~ \right],
\ee
where
\be
\label{VSS}
  V(t,r)  =  1 + \frac{1}{4}k(t)r^2~,
\ee
and $(\ldots)^{\centerdot} \equiv \partial/\partial t$. The function $a(t)$
plays the role of a generalized scale factor, $k(t)$ has the meaning of
a time-dependent ``curvature index", $r$ is the radial coordinate, and $c_0$ is the (constant) speed of light. The Stephani universe is complementary (and so very different) to Lemaitre-Tolman-Bondi (LTB) universe \cite{LTB}  and should not be mistaken with the latter one.

The energy density and pressure are given by
\begin{eqnarray}
\label{rhost}
\varrho(t) & = & \frac{3}{8\pi G} \left[ \frac{\dot{a}^2(t)}{a^2(t)} + \frac{k(t)c_0^2}{a^2(t)}
\right],\\
\label{pst}
p(t,r) & = & w_{eff}(t,r) \varrho(t) c_0^2 \\
&\equiv& \left[-1~~+~\frac{1}{3} \frac{\dot{\varrho}(t)}{\varrho(t)} \frac{ \left[ \frac{V(t,r)}{a(t)} \right]}
  { \left[ \frac{V(t,r)}{a(t)} \right]^{\centerdot}} \right]  \varrho(t) c_0^2, \nonumber
\end{eqnarray}
and generalize the standard Einstein-Friedmann equations into inhomogeneous models. The radial dependence of the effective barotropic index $w_{eff}(r,t)$ is due to the radial dependence of the fluid pressure and means that a comoving observer does not follow a geodesic. In fact, a comoving observer has a four-velocity with a non vanishing radial component and move in the radial direction in addition to its movement due to the expansion. Extra radial force pushes him out of a geodesic.

In this paper we will study the model (\ref{STMET}) \cite{JMP93,dabrowski95,PRD15} with $k(t) = \beta a(t)$ and
$\beta=\, const.$, which has a simplified metric
\be
\label{met}
ds^2 = - \frac{c_0^2}{V^2} dt^2 + \frac{a^2(t)}{V^2} \left( dr^2 + r^2 d \Omega^2 \right) .
\ee
This metric can be considered as defining spatially dependent effective speed of light $c(t,r) = c_0/V(t,r)$ (provided we work in a special frame in which the Einstein field equations (\ref{rhost})-(\ref{pst}) are valid \cite{VSL_theory}) or still can mimic the spatial dependence of the speed of light provided we take $c_0 \to c=c(t)$ in (\ref{met}) and make an appropriate ansatz (cf. our formula (\ref{c_ansatz})). Whether variability of $c$ is a reasonable idea seems still to be the matter of a debate with that whole spectrum views on the topic \cite{VSL_controversy}. Here we merely take a pragmatic position and try to study what would be the consequences of such an assumption in the context of inhomogeneity leaving the dispute open.

The definition of redshift for the Stephani model \cite{PRD15}  (which now depends on the radial coordinate as well) reads as
\be
\label{rs}
1+z=\frac{a_0}{a_{e}}\frac{V_e}{V_0}~~,
\ee
where index ``0'' refers to the present moment of time and index ``$e$'' to the time of emission of a signal. If an observer is placed at the center of symmetry, then $V_0 = 1$ (cf. an off-center observer discussion in Ref. \cite{off-center14}). The radial distance $r$ can be calculated from the condition of taking the null geodesic $ds^2 = 0$ in (\ref{met}) (replacing
$c_0 \to c=c(t)$ \cite{PLB14,JCAP14}), i.e.
\be
\label{r}
r = \int_{t_e}^{t_0} \frac{c(t)dt}{a(t)} .
\ee
Using the definition of dimensionless density $\Omega_{\beta}$:
\begin{equation}
\Omega_{\beta} = - \frac{\beta\, c^{2}(t)}{a(t)\, H^{2}(t)} \quad \Rightarrow \quad \Omega_{\beta,0} = - \frac{\beta\, c^{2}_{0}}{a_{0}\, H^{2}(0)} \; ,
\end{equation}
for inhomogeneity being adopted into the homogeneous (``averaged'') Friedmann equation, one can convert the redshift into
\begin{equation}
1+z = \frac{1}{a} - \frac{\Omega_{\beta,0}}{4} r(a)^{2}\; ,
\end{equation}
and so the Friedmann equation reads as
\begin{eqnarray}
H^{2}(a) &=& \frac{8\pi G}{3} \rho_{i} - \frac{k(t)}{a^{2}(t)} = \frac{8\pi G}{3} \rho_{i} - \frac{\beta}{a(t)} = \\
&=& H^{2}_{0} \left[ \frac{\Omega_{r,0}}{a^{4}} + \frac{\Omega_{m,0}}{a^{3+3(1+w)}} + \frac{\Omega_{\beta,0}}{a} f_{\beta}(a) \right]\; , \nonumber
\end{eqnarray}
with $\Omega_{m,0} = 1- \Omega_{r,0}-\Omega_{\beta,0}$ ($\rho_i$ labels the non-interacting fluids, $r$ is for radiation, and $m$ is for matter).

Here $f_{\beta}(a)$ depends on the ansatz for varying speed of light (VSL) theory. Given that a certain level of arbitrariness is intrinsic to VSL about the choice of a functional form for $c(t)$, we have focussed on three ans\"{a}tze, which we have dubbed as: a standard no-varying $c$ ansatz, Barrow-Magueijo ansatz \cite{VSL_theory}, an inhomogeneous ansatz respectively:
\begin{equation} \label{c_ansatz}
f_{\beta}(a) =\begin{cases}
1  \quad &\mbox{for } c(t) = c_{0} = const.  \\
a^{2n}(t)  &\mbox{for } c(t) = c_{0} a^{n}(t) \\
\frac{1}{V^{2}(t,r)} &\mbox{for } c(t,r) \equiv \frac{c_{0}}{V(t,r)}
\end{cases}
\end{equation}
with $V$ from (\ref{VSS}) now expressed as:
\begin{equation}
V(t,r) = 1-\frac{\Omega_{\beta,0}}{4}\, a \, r^2(a)\; .
\end{equation}
Now, $r(a)$ is the dimensionless comoving distance given by:
\begin{itemize}
 \item for $c(t) = c_{0} = const.$:
       \begin{equation}
       \label{stand}
       r(a) = \int^{1}_{x} \frac{1}{\sqrt{\Omega_{r,0} + \Omega_{m,0} a^{4-3(1+w)} + \Omega_{\beta,0} a^{3}}}
       \end{equation}
 \item for $c(t) = c_{0} a(t)^{n}$:
       \begin{equation}
       \label{ansatzBM}
       \int^{1}_{x} \frac{a^{n}}{\sqrt{\Omega_{r,0} + \Omega_{m,0} a^{4-3(1+w)} + \Omega_{\beta,0} a^{2n+3}}}
       \end{equation}
 \item for $c(t,r) = c_{0} / V(r,t)$:
       \begin{equation}
       \label{ouransatz}
       \int^{1}_{x} \frac{1-\Omega_{\beta,0}/4\, a\, r^2(a)}{\sqrt{\Omega_{r,0} + \Omega_{m,0} a^{4-3(1+w)} + \Omega_{\beta,0} \frac{a^{3}}{(1-\Omega_{\beta,0}/4\, a\, r^2(a))^2}}}
       \end{equation}
\end{itemize}
Such choices for the VSL analytic expressions are in some way strategic, because they stand for three different ways for VSL and inhomogeneity to be entangled: in the standard (classical) case, of course, we have no VSL, but only inhomogeneity; in the Barrow-Magueijo case, we assume a different temporal behavior for both VSL and inhomogeneity (which is also spatially dependent); while in the last ``inhomogeneous'' ansatz, we assume that the time variation of VSL is intrinsically correlated to the inhomogeneity.

\section{Angular diameter distance $D_A$ maximum in inhomogeneous universe}
\label{Angdiam}

The angular diameter distance for the Stephani model (\ref{met}) reads as \cite{SJ2006}
\be
\label{DA}
D_A =\frac{a(t)}{V(t,r)} r = \frac{a_0}{V_0(1+z)} r~.
\ee
where we have used the redshift definition (\ref{rs}).

After combining (\ref{DA}) and (\ref{r}) we are able to calculate the time derivative of $D_A$ (which is equivalent to taking $z$ derivative  since $z$ is monotonic with time here) and the condition for angular diameter distance maximum as
\be
\label{DAder}
\frac{\partial{D_A}}{\partial t} = \frac{\dot{a} V - \dot{V} a}{V^2} \int_{t_e}^{t_0} \frac{c(t)dt}{a(t)} - \frac{c}{V} = 0 .
\ee
This gives the relation which can be used to evaluate the timely and spatial dependence of the speed of light
\be
c(t,r) = D_A \left( H V - \dot{V} \right) ,
\ee
or
\be
D_A(t,r) = \frac{c(t,r)}{H V - \dot{V}} ,
\ee
which allows to relate the inhomogeneity with the variability of the speed of light $c$. In other words, variability of $c$ can be mimicked by spatial inhomogeneity, and vice versa, the inhomogeneity can be mimicked by the variability of $c$ (cf. our introductory discussion of an $\alpha$-dipole \cite{alphadipole} and the relation $\alpha \propto 1/c$).

Using (\ref{met}) and (\ref{r}), one has:
\be
\label{D_A}
D_A = \frac{c}{H + \frac{\beta}{2} c r} ,
\ee
which can be solved for $c$ as follows
\be
\label{DAc}
c = \frac{D_A H}{1 - \frac{\beta}{2} c D_A r} ,
\ee
or
\be
c = \frac{D_A H}{1 - \frac{\beta}{2}{D_A^2 \frac{V_0(1+z)}{a_0}}}
\ee
Here $V_0$ and $a_0$ are known (both can be taken equal to 1), then we are left with the need to have $D_A,\,H$ and $z$, while $\beta$ can be taken from the limits found in \cite{PRD15}, which are based on independent tests. Then, one may evaluate both: the variability of $c$ for a given inhomogeneity $\beta$ and the inhomogeneity $\beta$ assuming that c is not varying at all.

The expression for the maximum in the angular diameter distance can be finally written down from (\ref{DAc}) as:
\begin{equation}
c(a) = \frac{D_{A}(a){H(a)}}{1+\frac{\Omega_{\beta,0}}{2}\, a \, r^2(a)} \;,
\end{equation}
and so we have:
\begin{eqnarray}
\label{ans1}
c_{0} &=& \frac{D_{A}(a){H(a)}}{1+\frac{\Omega_{\beta,0}}{2}\, a \, r^2(a)} \; \mathrm{for} \; c(t) = c_{0} = const. ,
 \\
 \label{ans2}
c_{0} a^{n} &=& \frac{D_{A}(a){H(a)}}{1+\frac{\Omega_{\beta,0}}{2}\, a \, r^2(a)} \; \mathrm{for} \; c(t) = c_{0} a^n(t) ,
\\
\label{ans3}
\frac{c_{0}}{V(t,r)} &=& \frac{D_{A}(a){H(a)}}{1+\frac{\Omega_{\beta,0}}{2}\, a \, r^2(a)} \; \mathrm{for} \; c(t,r) = \frac{c_{0}}{V(r,t)} .
\end{eqnarray}
We implicitly assume that the relations (\ref{ans1})-(\ref{ans3}) are evaluated at the maximum $a = a_{M}$. In \cite{PRL15} we found that for homogeneous models we have:
\begin{equation}\label{eq:dahc}
D_{A}(a)H(a) = c(a)\; ,
\end{equation}
but with the assumption of no spatial curvature. In \citep{PRD16}, we show that this relation is valid, to some order, even for $k\neq0$, because contributions derived from present bounds on curvature are $\sim2$ order smaller than a VSL signal. Clearly, in a standard scenario of constant speed of light, this relation converts in:
\begin{equation}
\frac{D_{A}(a)H(a)}{c_{0}} = 1 .
\end{equation}
Instead, here we are considering curvature since the beginning, and the maximum relation is changed to:
\begin{eqnarray}
\label{Deltac}
 \Delta_c &=& \frac{D_{A}(a){H(a)}}{c_{0}} \\
&=& \begin{cases}
1+\frac{\Omega_{\beta,0}}{2}\, a \, r^2(a) & \mbox{for} \, c(t) = const. \\
a^{n} \left( 1+\frac{\Omega_{\beta,0}}{2}\, a \, r^2(a)\right) & \mbox{for} \, c(t) = c_{0} a^n(t) \\
\left[ \frac{1+\frac{\Omega_{\beta,0}}{2}\, a \, r^2(a)}{1- \frac{\Omega_{\beta,0}}{4} \, a \, r^2(a)} \right] & \mbox{for} \, c(t,r) = \frac{c_{0}}{V(r,t)} \nonumber
\end{cases}
\end{eqnarray}
Interestingly, we can note that even in the first case, with constant speed of light, the inhomogeneity might play the role of an ``effective'' VSL, in particular, inhomogeneity might mimic a time and space varying speed of light. Thus, in this case, if there were no spatial dependence, we would end with the same results of \citep{PRL15,PRD16}, where a VSL plus homogeneity was assumed. For the other two ans\"{a}tze, it would be actually impossible to discriminate between a pure VSL signal and a pure inhomogeneity, because the two are strongly coupled.

\section{Data Analysis}
\label{data}

The analysis involved a set of cosmological data which includes the Type Ia Supernovae (SNeIa), Barion Acoustic Oscillations (BAO), Cosmic Microwave Data (CMB) and a prior on the Hubble constant parameter, $H_0$.

\subsection{Type Ia Supernovae}
In our analysis we used the SNeIa (Supernovae Type Ia) data from the SCP (Supernova Cosmology Project) compilation \cite{Union2}. The $\chi^2_{SN}$ is defined as
\begin{equation}
\chi^2_{SN} = \Delta \boldsymbol{\mathcal{F}}^{SN} \; \cdot \; \mathbf{C}^{-1}_{SN} \; \cdot \; \Delta  \boldsymbol{\mathcal{F}}^{SN} \; ,
\end{equation}
with $\Delta\boldsymbol{\mathcal{F}}^{SN} = \mathcal{F}^{SN}_{theo} - \mathcal{F}^{SN}_{obs}$ being the diffrence between the theoretical and the observed value of the observable quantity ${\mathcal{F}}^{SN}$ and $\mathbf{C}_{SN}$ is the covariance matrix. For the SCP, the observed quantity will be the predicted distance modulus $\mu$ of the SNeIa which reads
\begin{equation}
\mu(z)=5\log_{10}d_L(z)+25
\end{equation}
where $d_L$ is the luminosity distance in our Stephani model \cite{stelmach01,GSS}:
\begin{equation}
d_L=(1+z) a_0 r(a)
\end{equation}
with $r(a)$ given by $(\ref{stand})$, $(\ref{ansatzBM})$ or $(\ref{ouransatz})$.

\subsection{Baryon Acoustic Oscillations}

The $\chi^2_{BAO}$ for Baryon Acoustic Oscillations (BAO) is given by
\begin{equation}
\chi^2_{BAO} = \Delta \boldsymbol{\mathcal{F}}^{BAO} \; \cdot \; \mathbf{C}^{-1}_{BAO} \; \cdot \; \Delta  \boldsymbol{\mathcal{F}}^{BAO} \; ,
\end{equation}
where  $\mathcal{F}^{BAO}$ is a vector quantity composed of the values of the two following quantities: the acoustic parameter
\begin{equation}\label{eq:AWiggle}
A(z) \equiv  \sqrt{\Omega_{m}} H_0 \frac{D_{V}(z)}{c(z) \, z} \, ,
\end{equation}
and the Alcock-Paczynski distortion parameter
\begin{equation}\label{eq:FWiggle}
F(z)  \equiv (1+z)  \frac{D_{A}(z)\, H(z)}{c(z)} \, ,
\end{equation}
both evaluated in the WiggleZ Dark Energy Survey \citep{WiggleZ_0} at redshifts $z=\{0.44,0.6,0.73\}$ and their values are given in Table~1 of \citep{WiggleZ}. The quantities $D_A$ and $D_V$ which occur in (\ref{eq:AWiggle}) and (\ref{eq:FWiggle}) are respectively the angular diameter distance given by (\ref{DA}) and the volume distance defined as
\begin{eqnarray}\label{eq:dV}
\nonumber
D_{V}(z)  &=& \left[ (1+z)^2 D^{2}_{A}(z) \frac{c(z) \, z}{H(z)}\right]^{1/3}\\
&=& \left[ a_0^2 r^2(a) \frac{c(z) \, z}{H(z)}\right]^{1/3}.
\end{eqnarray}
where $c(z)$ is the value of the speed of light at redshift z given by one of the three ans\"{a}tze introduced in formula $(\ref{c_ansatz})$.

\subsection{Shift parameter}
The position of the Cosmic Microwave Background (CMB) acoustic peaks depends on the geometry of the considered model and, as such,
can be used to discriminate between dark energy models of the different nature. The quantity that we will use here is  the so-called shift parameter  defined as
\begin{equation}
R\equiv \sqrt{\Omega_m H^2_{0}} \frac{r(z_{\ast})}{c(z_{\ast})}~,
\end{equation}
where $r(z_{\ast})$ is the comoving distance evaluated at the photon-decoupling redshift $z_{\ast}$ given by the fitting formula \citep{Hu}
\begin{equation}{\label{eq:zdecoupl}}
z_{\ast} = 1048 \left[ 1 + 0.00124 (\Omega_{b} h^{2})^{-0.738}\right] \left(1+g_{1} (\Omega_{m} h^{2})^{g_{2}} \right) \, ,
\end{equation}
with
\begin{eqnarray}
g_{1} &=& \frac{0.0783 (\Omega_{b} h^{2})^{-0.238}}{1+39.5(\Omega_{b} h^{2})^{-0.763}}
\end{eqnarray}
and
\begin{eqnarray}
g_{2} &=& \frac{0.560}{1+21.1(\Omega_{b} h^{2})^{1.81}}~,
\end{eqnarray}
where the parameters $\Omega_b h^2$ and $\Omega_m h^2$ represent  the physical baryon and dark matter density of the $\Lambda$CDM model respectively.
The $\chi^2_{R}$ for the CMB shift parameter is \cite{WangDai}
\begin{equation}
\chi^2_{R}=\frac{({\cal R}-1.7482)^2}{0.0048^2}~.
\end{equation}

For the purpose of our analysis we have assumed a gaussian prior on the Hubble constant, $H_0$ \citep{Bennett}:
\begin{equation}
\chi^{2}_{H_{0}} = \frac{(H_{0}- 69.6)^{2}}{0.07^2}~.
\end{equation}
Thus, the total $\chi^{2}$ will be the sum of: $\chi^{2}_{SN},\chi^{2}_{BAO},\chi^{2}_{R},\chi^{2}_{H_{0}}$. We minimize $\chi^{2}$ using the Markov Chain Monte Carlo (MCMC) method.

\section{Results and Conclusions}
\label{results}

The results from the MCMC can not only be used to infer the statistical properties of our models, but also directly to calculate the redshift location of the maximum in the angular diameter distance, and the amount of deviation from constant $c$ which is expected and which is compatible with observations.

{\renewcommand{\tabcolsep}{1.5mm}
{\renewcommand{\arraystretch}{2.}
\begin{table*}[htbp]
\begin{minipage}{0.85\textwidth}
\caption{Results.}
\label{Table1}
\centering
\resizebox*{\textwidth}{!}{
\begin{tabular}{c|cccc|cc}
\hline \hline
 & $H_{0}$ & $\Omega_{\beta}$ & $w$ & $n$ & $z_{M}$ & $\Delta_{c}$ \\
\hline
$c(t) = c_{0} = const.$   & $69.6^{+0.7}_{-0.7}$ & $0.682^{+0.022}_{-0.023}$ & $-0.014^{+0.004}_{-0.004}$ & $-$ & $1.553\pm0.026$ & $1.140\pm0.011$ \\
$c(t) = c_{0} a^{n}(t)$   & $69.6^{+0.7}_{-0.6}$ & $0.638^{+0.031}_{-0.029}$ & $-0.139^{+0.047}_{-0.045}$ & $-0.083^{+0.034}_{-0.034}$ & $1.816\pm0.132$ & $1.281\pm0.074$ \\
$c(t,r) = c_{0} / V(r,t)$ & $69.6^{+0.7}_{-0.7}$ & $0.669^{+0.022}_{-0.022}$ & $0.003^{+0.003}_{-0.003}$ & $-$ & $1.708\pm0.042$ & $1.200\pm0.015$ \\
\hline \hline
\end{tabular}}
\end{minipage}
\end{table*}


The best-fit parameters for inhomogeneous models $\Omega_{\beta}, w, n$ obtained for the data given in Section \ref{data} are listed in the left panel of the Table \ref{Table1}. Using the values derived one was able to evaluate the maximum redshift $z_M$.
Following \cite{PRL15,PRD16} we have considered the CPL \citep{CPL} $w+w_{a}$ \texttt{plikHM}$\_$\texttt{TTTEEE}$\_$\texttt{lowTEB}$\_$\texttt{BAO}$\_$\texttt{post}$\_$\texttt{lensing} and the baseline ($\Lambda$CDM) model
\texttt{plikHM}$\_$\texttt{TTTEEE}$\_$\texttt{lowTEB}$\_$\texttt{lensing}$\_$\texttt{po\\st}$\_$\texttt{BAO}$\_$\texttt{H070p6}$\_$\texttt{JLA} bestfit from the \textit{Planck 2015} release. We have taken into account $10^4$ cosmological models, derived from varying the cosmological parameters consistently within the $1\sigma$ confidence intervals defined for the previous parametrization. As pointed out in \citep{PRL15,PRD16}, of course, the CPL parametrization is only one of the many dark energy phenomenological models available, but it is somewhat used as a reference model in the literature. Moreover, the large errors on its parameters, in particular on the dynamical dark energy EoS parameter $w_a$, make us confident on having explored a very large set of cosmological scenarios compatible with observational data, thus making our estimation for the range of $z_M$ highly conservative. For this reason, we also consider a much more restrictive cosmological constant case (baseline model) which is recognized, at the preset stage of observations, as the best consensus cosmological model.

At the end, it results that for the CPL case, $z_{M}$ lies in the range $[1.4,1.75]$ for more than $99\%$ of $10^{4}$ random cosmological models chosen as described above, while for the $\Lambda$CDM case, $z_{M}$ lies in the range $[1.57,1.62]$. The values we found for the inhomogeneous model are more or less compatible with these ranges; given the distribution of the $10^{4}$ models, we have calculated what is the probability to find the maximum in the ranges in Table~\ref{Table1} $(1\sigma$ confidence intervals), and we have: $29\%$ for $c$ constant; $5\%$ for the time-varying-only $c$; $8\%$ for the space-time-varying $c$, when compared to the CPL case. Thus, given the bounds from present data, we may conclude that, while the maximum detection for the first ansatz would not constitute, by its own, a net statistically significant proof for a possible VSL and/or inhomogeneity signal to be detected, in the other cases it would give a stronger hint for a breaking of the $c$-constancy and cosmological background homogeneity which are at the base of the standard cosmological scenario. Anyway, it is plain that simply using the maximum detection, we would not be able to discriminate between a classical (constant $c$, homogeneous background and a time-varying dark energy) scenario and an alternative one in a fully valid statistical way. Instead, when comparing our results in Table~\ref{Table1} with the $\Lambda$CDM case, in the first case, the model would be ruled out at almost $3\sigma$; while the second and third ans\"{a}tze would be completely discarded.

On the other hand, the location of the maximum would not be the only probe we have to test our hypothesis. In fact, if we use the definition of $\Delta_{c}(a)$ given by (\ref{Deltac}), then we can calculate the deviation from constant speed of light $(c_{0})$ measured at the maximum. Even if the maximum location should not be strikingly different from the classical scenario (as it is the case of the first ansatz), still we should refer to the value of $\Delta_c$, if it was equal to one, or not. For all our ans\"{a}tze, such deviation is presented in the last column of Table~\ref{Table1}. As we pointed out in \citep{PRD16}, in principle, SKA will be able to detect a $1\%$ deviation from constant speed of light at $3\sigma$ confidence level at the maximum redshift. It is clear that all the models we have considered here, exhibit variations which are fully detectable, being of the order of $10\%$. Thus, our models have the merit of being completely falsifiable. If no signal of such order of magnitude will be detected, it will be a clear signature of no inhomogeneity at play; still a VSL might be possible, but not a spatial inhomogeneity.

\section*{Acknowledgements}

This paper was financed by the Polish National Science Center Grant DEC-2012/06/A/ST2/00395.

\vfill
\end{document}